\documentclass[journal,transmag]{IEEEtran}
\usepackage[utf8]{inputenc}
\usepackage{CJKutf8}
\usepackage{threeparttable,placeins,makecell,stfloats,cite}
\usepackage{threeparttable,arydshln}
\usepackage{amsmath,amssymb,bm}
\usepackage[dvips]{graphicx}
\usepackage[usenames]{color}
\usepackage{amsfonts}
\usepackage{latexsym}
\usepackage{multirow}
\usepackage{rotating}
\usepackage{float}
\usepackage[format=hang]{subfig}
\usepackage{url}
\usepackage{cite}
\usepackage{algorithmic}
\usepackage{algorithm}
\usepackage{diagbox}

\begin{document}

\title{ A Dynamic YOLO-Based Sequence-Matching Model for Efficient Coverless Image Steganography}

\author{\IEEEauthorblockN{Jiajun Liu\IEEEauthorrefmark{1}, 
Lina Tan\IEEEauthorrefmark{1,*},
Zhili Zhou\IEEEauthorrefmark{2},
Yi Li\IEEEauthorrefmark{1},
Peng Chen\IEEEauthorrefmark{1}}
\IEEEauthorblockA{\IEEEauthorrefmark{1}School of Computer Science at Hunan University of Technology and Business, Changsha 410205, China}
\IEEEauthorblockA{\IEEEauthorrefmark{2}Institute of Artificial Intelligence, Guangzhou University, China}}

\maketitle

\begin{abstract}
Many existing coverless steganography methods establish a mapping relationship between cover images and hidden data. There exists an issue that the number of images stored in the database grows exponentially as the steganographic capacity rises. The need for a high steganographic capacity makes it challenging to build an image database. To improve the image library utilization and anti-attack capability of the steganography system, we present an efficient coverless scheme based on dynamically matched substrings. 
YOLO is employed for selecting optimal objects, and a mapping dictionary is established between these objects and scrambling factors.  With the aid of this dictionary, each image is effectively assigned to a specific scrambling factor, which is used to scramble the receiver's sequence key. To achieve sufficient steganography capability based on a limited image library, all substrings of the scrambled sequences hold the potential to hide data. After completing the secret information matching, the ideal number of stego images will be obtained from the database. According to experimental results, this technology outperforms most previous works on data load, transmission security, and hiding capacity. Under typical geometric attacks, it can recover 79.85\% of secret information on average. Furthermore, only approximately 200 random images are needed to meet a capacity of 19 bits per image.

\end{abstract}


\section{Introduction}
Steganography can transmit secret information through various media such as videos, images, audio, or texts. Even though image steganography has advanced significantly \cite{2008Adaptive,2007Strange,2017A,2002Secure,2008An}, modifications to image pixels or transform domain coefficients always result in changes to the statistical characteristics of the images, which may still be recognized by some steganalyzers.

Coverless steganography \cite{2015Coverless} directly selects images as stego ones based on mapping rules. Zheng et al. \cite{2017SIFT} enhanced the method with SIFT feature points, improving resistance to rotation and scaling attacks. Zou et al. \cite{2019A} introduced a model based on average pixel values for increased steganography capacity. In \cite{2018Robust, 2019Coverless}, sub-block coefficients in the transform domain are used for robust feature sequences, which have improved resistance to noise attacks compared to previous models, but are sensitive to geometric attacks.

Recently, deep learning models have been increasingly integrated into coverless steganography based on mapping rules. In 2020, Luo et al. \cite{2020Coverless} proposed a Faster RCNN-based steganography, establishing a mapping dictionary between objects and labels for secret information concealment. This method is robust against geometric attacks but less tolerant to noise compared to previous works.
In the same year, Liu et al. \cite{2020DenseNet} employed DenseNet \cite{2016Densely} to extract high-dimensional CNN features for hash sequence mapping. Another approach \cite{2020Segment} hides secret information by semantic feature extraction and image object region segmentation.
Additionally, a coverless scheme using camouflage images and CNN features was introduced \cite{Camouflage}, which has more flexible capacity settings and robustness against image attacks. However, the above techniques usually require building a sizable image database, posing challenges in maintaining uniqueness and preventing feature collisions amongst those extracted from different images. 

Cover image generation is an alternate way of coverless steganography that uses generative models, such as generative adversarial networks (GAN) \cite{GAN}, to produce stego images for the purpose of hiding secret information.
An image texture synthesis technique based on ACGAN was proposed in 2017 \cite{2017Coverless}, wherein the hidden data are transformed into noise and fed into the ACGAN network. However, there was a lack of realism in the generated images. Zhang et al. \cite{2019An} used GAN to generate multiple texture images, which then formed a mapping to binary sequences. However, diversity breaks down when the number of produced images reaches a threshold.
To increase the hiding capacity, Chen et al.\cite{2020A} applied SIFT \cite{2004Distinctive} to select images and StarGAN \cite{0StarGAN} to generate new images based on mapping face attribute features to hidden data. Although the security and robustness are achieved, the capacity remains limited.
A steganography method based on multi-domain image transformation was proposed\cite{multi-domain} to address the issue of insufficient cover images. The secret message is hidden using a generator and recovered using a classifier.
Inspired by irreversibility issues with traditional neural networks, a method \cite{GradientDescent} iteratively updates the noise vector using gradient descent for data extraction. While these methods  lessen the load on the image database, the resulting images lack quality and diversity, making it easier for attackers to detect variations. Moreover, the image content may be lost or altered during network operations, hindering accurate data extraction.

For the idea of coverless steganography, we consider making the length of the information hidden in each cover image variable rather than fixed, in order to improve the hiding capability while reducing the burden on the database. A sequence of length $u$ generates $(u^2+u)/2$ binary substrings, representing an image as such. This enables dynamic matching of secret information fragments with substrings of varying lengths. The larger the $u$ is, the more substrings are yielded for information matching, and relatively fewer images are needed in the database.
To address coverless steganography's limited robustness, we leverage high-level semantic features for flexible mapping rules. Recent advancements in object detection models, especially YOLO, have shown superior real-time performance and high average accuracy compared to techniques like DPM and R-CNN. This solution uses YOLO mainly due to its fast inference speed and stable feature extraction. The main contributions of this paper are as follows:
\begin{enumerate}
\item In view of the superior detection accuracy and inference time of YOLOs, we explored an Optimal Object Filtering Algorithm (OOFA) using YOLO-v5 to establish a robust mapping rule between the secret information and cover images. Tests conducted under geometric and noise attacks demonstrate the remarkable robustness of our method, which reaches nearly 87.4\%.
\item To increase the hidden bits per image, we extend the sequence key length sufficient to dynamically match the secret information with multiple substrings. Experiments show that, under ideal conditions, our approach outperforms existing methods with an average hiding capacity of 19 bits per image. 
\item To mitigate the impact of hiding capacity on image database load, we hide the information indirectly in the keys rather than in the cover images. In contrast to other approaches,  ours requires fewer cover images to meet the same hiding capacity.
\end{enumerate}

\section{The proposed method}
\subsection{Overall Framework}
\begin{figure*}[htpb]
	\centering
	\includegraphics[height = 7.5cm, width=15cm]{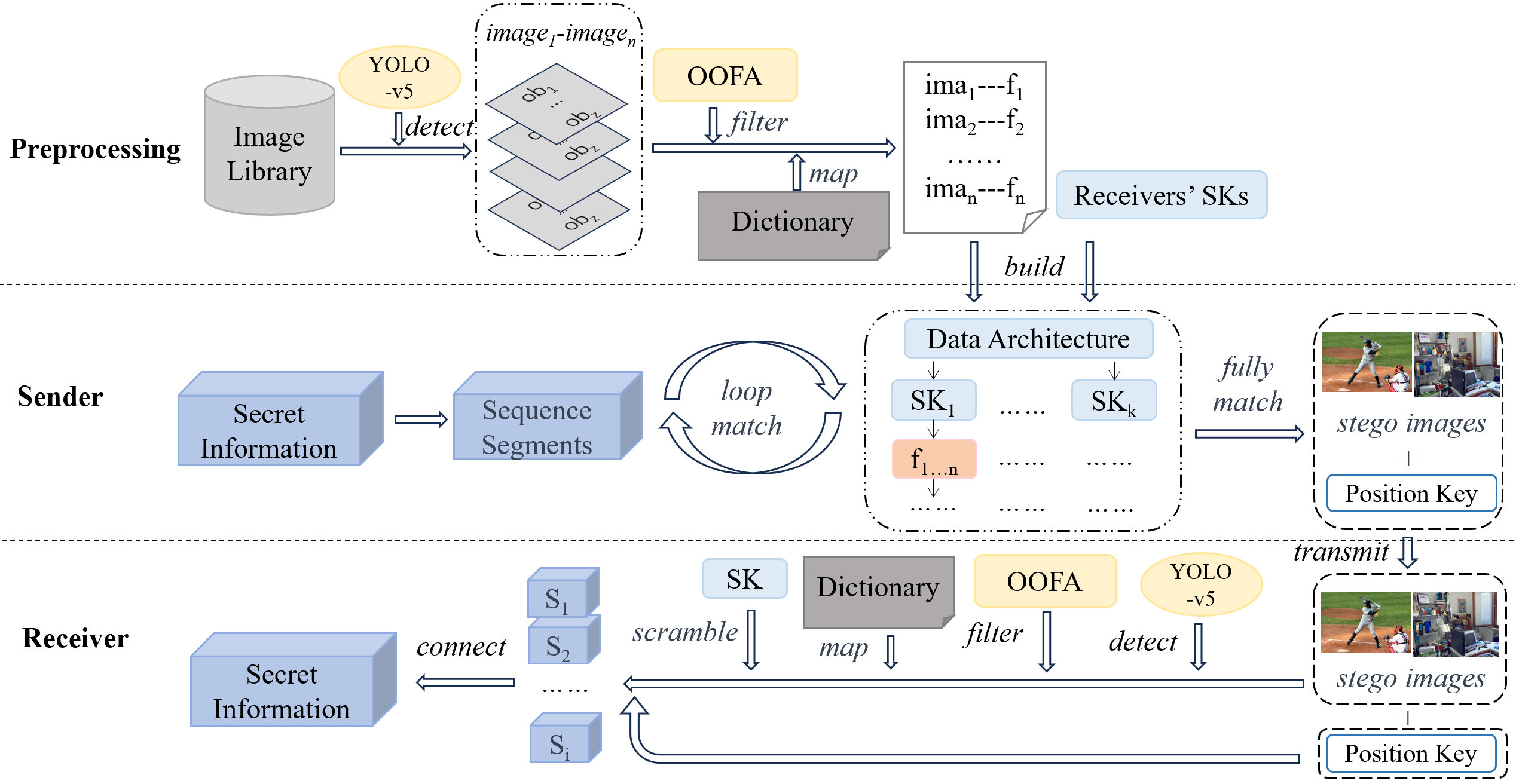} \\
	\caption{ Flowchart of The Coverless Image Steganography Algorithm}
	\label{framework}
\end{figure*}

Fig. \ref{framework} display the suggested framework. Its three main components are preprocessing, information hiding, and information extraction. During the preprocessing stage, each receiver has a unique sequence key. Once YOLO identifies all cover images in the database, a mapping dictionary must be established to generate the scrambling factors. These factors will then be used to scrambled the sequence keys, leading to the development of a data architecture based on the relationships between the sequence keys, scrambling factors, scrambled sequences, and cover images. In the stage of information hiding, after secret data matches substrings of scrambled sequences, the sender transmits the position key and corresponding stego images to the receiver. For each stego image, the receiver employs YOLO with the same settings to detect the optimal objects in it, and extract secret message segments through the sequence key, factors, and position key. 

\subsection{Preprocessing}
In our system, the sender and each receiver agree on a specific sequence key while ensuring perfect confidentiality. Preprocessing tasks like establishing the mapping dictionary, producing the scrambling factors, and creating the inverted index, are necessary before transferring the secret information.

\subsubsection{Mapping Dictionary Creation}
A mapping dictionary needs to be built to convert the object labels of images recognized by YOLO into scrambling factors. After all images in the database are detected by YOLO, a label list is obtained for all the objects. These labels are sorted alphabetically and each object will be assigned a unique scrambling factor according to the mapping dictionary. The mapping rules can be updated periodically to ensure security, such as in ascending or descending order.

\subsubsection{Sequence Key Distribution}
Each receiver must be issued a unique sequence key of length $t$ whose value is empirically set to 10000 following capacity and time cost testing. There are two approaches to assigning each receiver a specific sequence key. One way is to create a unique sequence by using the recipient ID as a random seed and reordering the initial sequence by the pseudo-random number generator. Another way is to manually distribute a bit-string of length $t$ to each recipient, which exhibits a high degree of randomness, and when combined with proper security measures, it becomes exceedingly challenging for any third party to deduce the assigned bit strings. Considering that the system is more vulnerable to attacks if the attacker is skilled with the pseudo-random function and receiver IDs, we choose the first method for key distribution.

\subsubsection{Scrambling Factor Generation}
The previously created mapping dictionary has established a mapping relationship between the image objects and scrambling factors. YOLO may find multiple objects from an image since it handles multi-object classification and localization. It’s worth noting that those objects with lower category probabilities may not be detected again after the image is attacked, which would result in poor robustness of the steganography model. Therefore, our algorithm will eliminate those undesirable bounding boxes and keep only the ones with the highest category probability. The generation process of scrambling factors is given as follows.

\begin{enumerate}
    \item Detect an input image $c$ with YOLO to get all the objects $OB$. The area of the bounding box and class probability of each object are recorded as $A(ob)$ and $P(ob)$ respectively.
    \begin{align}\label{1}
        OB=ob_{1}||ob_{2}||...||ob_{n}
    \end{align}
    \item In general, objects with larger area or higher category probability tend to represent better robustness. Therefore, we set a threshold $P$ and $A$ to filter out the objects with low robustness performance. After the objects whose bounding box area is less than $A$ are filtered out, the one with class probability exceeding $P$ and maximum is selected among the remaining objects.
\begin{align}
    OB_{filtered} = \{ob_{i} \in OB | A_i > A\} 
\end{align}
    \begin{align}\label{2}
       ob_{opt} = \mathop{\arg \max}\limits_{ob_i \in OB_{filtered}} \{P_i | P_i > P\} 
    \end{align}
    \item If the $ob_{opt}$ can not be found, it means that the image is not suitable as a stego image, then the image is supposed to be discarded, otherwise, a scrambling factor is generated according to the dictionary D.
    \begin{align}\label{3}
       f=
        \begin{cases}
        Null,& if\ ob_{opt} == Null; \\
    	D[ob_{opt}], & else; \\
	    \end{cases}
    \end{align}
\end{enumerate}

After the above steps, images with a suitable object can generate a scrambling factor. Through a lot of experiments, we believe that the value of $A$ is 15\% of the total pixels of the image is suitable. Moreover, we found that the class probability of some objects may not decrease after being attacked, or even increase, so we set the $P$ to 50\%.

\subsubsection{Building Data Architecture}
After the previous three steps, we can build a four-level data architecture. Since it is generated and saved in advance, it is unnecessary for the sender to reprocess all images before hiding the secret information, which can save vast amount of time for the information hiding process. The first level are the sequence keys owned by all receivers, the second level are the scrambling factors generated by the images, and the third level are the scrambled sequence obtained by scrambling sequence keys. Each fourth level contains a list of images, and the same scrambling factor can be extracted from the images in the same list. 

\begin{figure}[htbp]
	\centering
	\includegraphics[height = 3.2cm, width=8.5cm]{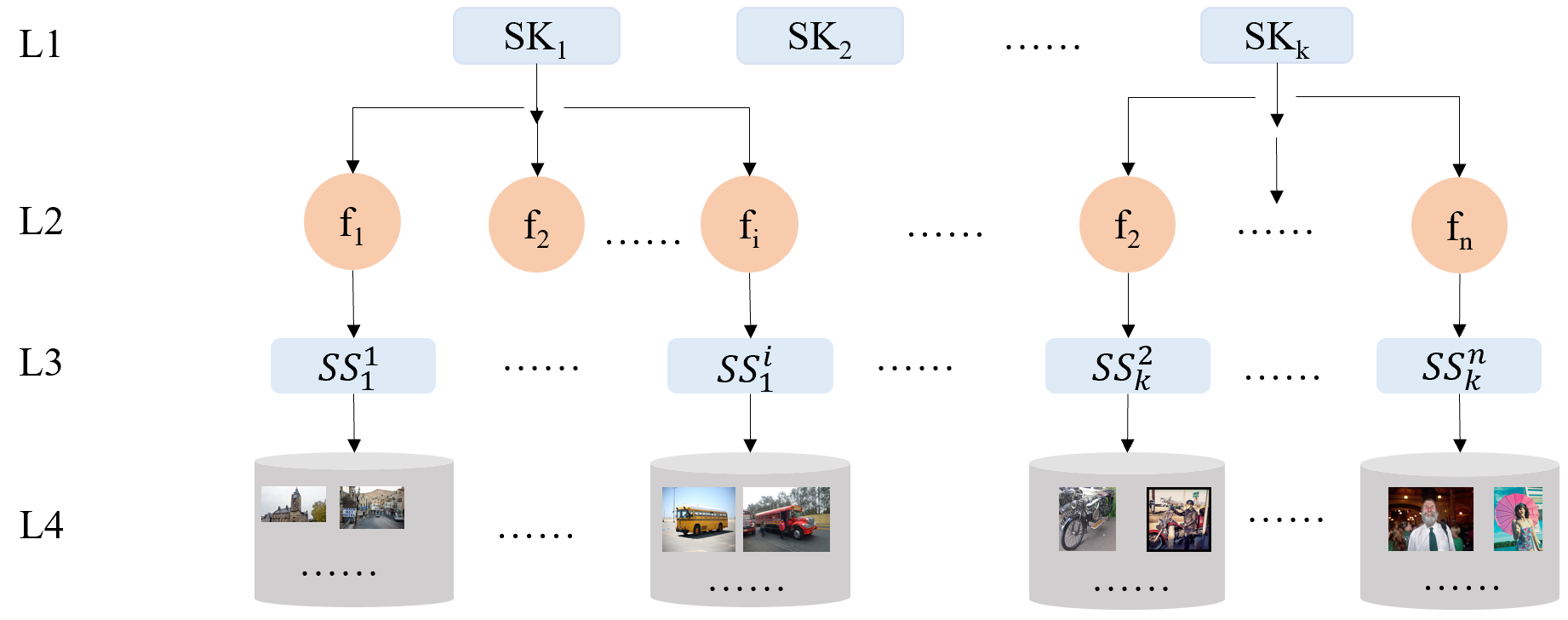} \\
	\caption{The Data Architecture of Image Database}
	\label{indexfigure}
\end{figure}

It is important to note that some images have the same suitable object, they will be mapped to the same scrambling factor, and these images will be putted into the same image list. Fig. \ref{indexfigure} shows the data architecture.

\subsection{Information Hiding}
Before selecting stego images, the secret information $SM$ needs to be converted into a binary stream $M$ whose length is denoted as $L$. The length of the sequence key also needs to be considered during the matching process, which is noted $t$. Then match the image in the data architecture mentioned above, which can be divided into the following steps:
\begin{enumerate}
    \item  For $MG$, select its first $n$ bits $MG=mg_1||mg_2||...||mg_n$. Due to the impossibility of matching extensive information within brief sequences, the value of $n$ in the first matching is computed by formula \ref{4} to steer clear of such futile matching endeavors.
\begin{equation}\label{4}
    n_{first} = min(t, L)
\end{equation}
    \item  Match all substrings with length $n$ in the third level of the data architecture , $Match(MG, SS)==0$ indicates the matching is failed, then set $n$ to $n-1$, repeat this step. Otherwise, a stego image $si_m$ could be obtained.
    \begin{align}
       \begin{cases}
	        n = n-1, & if \ Match(MG, SS)==0; \\
        	SI = SI || si_m,& else; \\
	   \end{cases}
    \end{align}
\begin{equation}
     Match(MG, SS) = \mathop{max}\limits_{i=1}^{L-n+1}\delta(MG, SS[i:i+n-1])
\end{equation}
In addition, the location information $key_m=\{k_f,k_l\}$ (including the left position index and length of the substring) should be record.
   \begin{equation}
         k_f=argmax(Match(MG, SS))
   \end{equation}
\begin{equation}
         k_l=n
   \end{equation}
    \item Regard the rest of information as a new piece of information, and set $L$ to $L-n$, repeat the previous steps until all the secret information is matched successfully, and get the stego images $SI$.
    \begin{align}\label{6}
       SI=si_1||si_2||...||si_m
    \end{align}

    \item Since the position and length of substrings are also the key factors to recover secret information accurately, the $Key$ should be transmitted to the receiver. We encrypt $Key$ prior to transmission using the AES algorithm out of consideration for security.
    \begin{align}\label{7}
       Key=key_1||key_2||...||key_m
    \end{align}
\end{enumerate}

This algorithm design ensures the security of secret information while maximizing the hiding capacity of the stego image. Furthrmore, we can pass $Key$ and stego images to the receiver respectively in different time periods and channels. 

\subsection{Information Extraction}
\begin{figure}[htbp]
	\centering
	\includegraphics[height = 4.4cm, width=8.5cm]{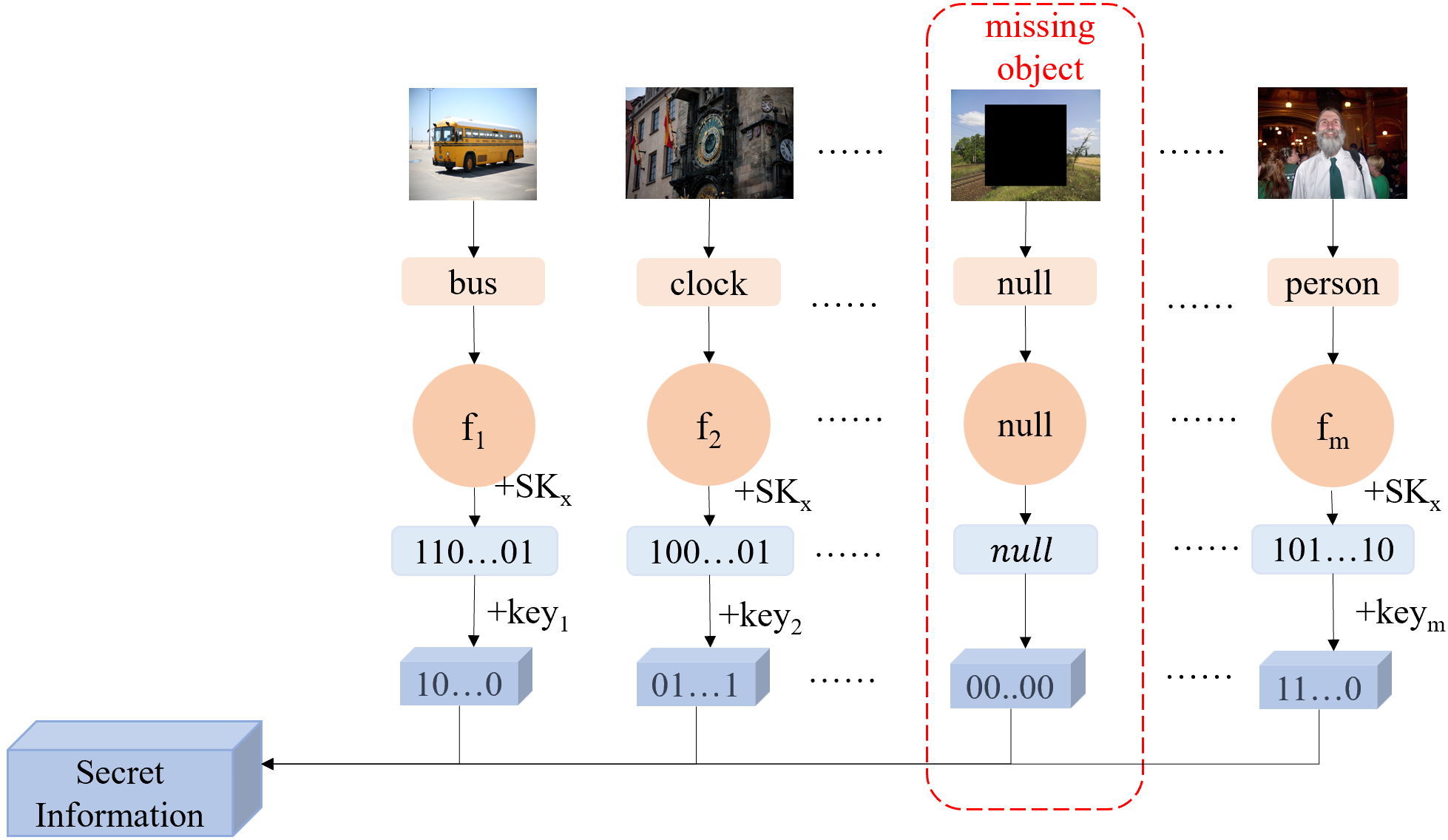} \\
	\caption{Process Chart for Information Extraction}
	\label{addzero}
\end{figure}

Some stego images may be attacked during transmission and the suitable objects fails to be detected, so that the receiver cannot restore the information accurately. For avoiding a missing object affecting the extraction of other location information, the the same number of '0' need to be added to the missing position. The $Key$ can provide us with the lengths of information hidden in stego images. This operation is shown in Fig. \ref{addzero}.

The steps of information extraction are as follows
\begin{enumerate}
    \item  Firstly, get all the objects $OB$ after using YOLO to detect the $j$th stego image $si_j$, with $1 \leq j \leq m$.
    \begin{align}\label{8}
       OB_j= ob_{1}||ob_{2}||...||ob_{n}
    \end{align}

    \item Secondly, the objects with small bounding box area less than $A$ are filtered out from $OB_j$, then The object with the highest classification probability is chosen as $ob_{opt}$.
\begin{align}
    OB_{filtered} = \{ob_{i} \in OB_j | A_i > A\} 
\end{align}
    \begin{align}\label{2}
       ob_{opt} = \mathop{\arg \max}\limits_{ob_i \in OB_{filtered}} \{P_i | P_i > P\} 
    \end{align}
    \item Match $ob_{opt}$ in the mapping dictionary $D$ to get a scrambling factor $f_j$. Then $f_j$ is used to scramble the sequence key $SK_x$ own by receiver to generate the scrambled sequence $SS_x^j$.
    \begin{align}\label{10}
       f_j=D[ob_{opt}]
    \end{align}
    \begin{align}\label{11}
       SS_x^j=Scrambling(SK_x,f_j)
    \end{align}

    \item Extract the hidden secret information $mg_j$ of the image from $SS_x$ through $key_j$.
    \begin{align}\label{12}
       mg_j=SS_x^j[key_j[0],key_j[0]+key_j[1]-1]
    \end{align}

    \item Finally, the all information fragments extracted from stego images are spliced together to recover the binary stream representation of secret information $M$, then $M$ is converted to secret information $SM$.
    \begin{align}\label{13}
       M = mg_1||mg_2||...||mg_m
    \end{align}

\end{enumerate}

\section{Experiments and Analysis}
Experimental environment: Intel(R) Xeon(R) Silver 4310 CPU @ 2.10GHz, 30.00GB RAM and one Nvidia GeForce RTX A4000 GPU. All experiments are completed in Pycharm and MATLAB R2021a.

Data set: MS COCO 2017 includes 118287 training images, 5000 validation images, and 40670 test images. This data set contains natural pictures and common object pictures in life and and is known for its complex background and relatively large number of objects, making it ideal for training YOLO-v5.

\subsection{Hiding Capacity}

\subsubsection{Capacity Test of Our Method}
The steganography capacity is influenced by the sequence key length and scrambling factors. We randomly selected images (100 to 5000) from the COCO dataset, forming six databases. Tests were conducted with sequence key lengths ranging from 100 to 50000 bits. Results in Table \ref{capacityours} indicate that beyond a certain image number, capacity impact is negligible. Within the same size database, steganography capacity increases with key length, but the increment diminishes. Considering retrieval efficiency, we recommend a 10000-bit key, yielding around 19 bits of capacity.

\begin{table*}[ht]
	\centering
	\caption{Hiding capacity of our scheme}
	\begin{tabular}{ccccccccccc}
		\hline
		\diagbox{images}{length(bits)}& 100 & 400 & 800 & 1000 & 2000 & 5000 & 8000 & 10000 & 15000 & 20000  \\
	    \hline
        100 & 11.688& 14.754& 14.876& 15.517& 16.364& 17.822& 18.182& 18.947& 19.149& 19.565\\
        500 & 12.676& 14.876& 15.652& 16.071& 17.143& 18.557& 19.149& 19.565& 20.0&   20.455\\
        1000& 12.676& 14.634& 15.929& 16.216& 17.143& 18.557& 19.355& 19.355& 20.225& 20.690 \\
        1500& 12.857& 15.0  & 16.216& 16.364& 17.308& 18.75 & 19.565& 19.565& 20.225& 20.455 \\
        2000& 12.950& 15.0  & 16.216& 16.822& 17.647& 18.75 & 19.565& 19.565& 20.455& 20.690 \\
        5000& 13.043& 15.0  & 16.364& 16.822& 17.876& 18.75 & 19.565& 19.565& 20.455& 20.930 \\
     \hline
	\end{tabular}
	\label{capacityours}
\end{table*}

\subsubsection{Comparison of Capacity}
In this comparison, our method is evaluated against others relying on mapping rules. Due to some unreproducible models, comparison data was extracted from the original papers or their citations. The comparison results are given in the Table \ref{capacitycomparision}. In particular, in \cite{2020Segment,2020Coverless,2019Faster}, the value of $n$ is $1\sim3$ in general. It is concluded that our method still has a slight advantage over these methods. 

\begin{table}[h]
	\centering
	\caption{The capacity of schemes}
	\begin{tabular}{cc}
		\hline
		methods & capacity (bits/image)\\
        \hline
        ours & 19\\
        CI-CIS \cite{Camouflage}& $1 \sim 15$ \\
        Mask RCNN \cite{2020Segment}& $(8 \sim 15)\times n$ \\
        Multiple Objects \cite{2020Coverless}& $ 6 \times n $\\
        Faster RCNN \cite{2019Faster}& $5 \times n$\\
        Pixel \cite{2015Coverless}& 8\\
        SIFT \cite{2017SIFT}& 8\\
        DCT \cite{2018Robust}& $1 \sim 15$\\
        DWT \cite{2019Coverless}& $1 \sim 15$\\
        \hline
	\end{tabular}
	\label{capacitycomparision}
\end{table}

\subsection{Database Size}
In order to carry out image matching on every conceivable binary stream of confidential information, it is essential to have a comprehensive database. However, the current coverless steganography is limited by the length of the hash, and the reason why the length of the hash cannot not be overly long is because of the difficulty in constructing a comprehensive database. Therefore, it is worth exploring how to effectively reduce the size of the image database.

\subsubsection{Database Size Test of Our Method}
In our setup, each object is linked to a unique scrambling factor, determined by the quantity of identified entities fulfilling the hidden demand. We explore the impact of different database sizes on scrambling factors, conducting experiments and evaluating their numbers. Notably, in Table \ref{capacityours}, a database size within the 1500 range significantly influences steganography capacity. We randomly selected images (50 to 1200) from the COCO dataset, conducted experiments, and determined the average scrambling factors per dataset. Table \ref{sizeours} shows a substantial increase in scrambling factors as the database size rises from 50 to 200. Beyond 200 images, increasing the database size to 1200 does not notably affect scrambling factors, indicating that 200 cover images suffice for steganography requirements. 
\begin{table}[htp]
	\centering
	\caption{Database size test(the length of sequence is 10000)}
	\begin{tabular}{ccc}
		\hline
		number of images & number of factors& capacity(bits/image)\\
		\hline
		 50  &  22 & 17.947\\
         100 &  34 & 18.680\\
         200 &  50 & 19.151\\
         400 &  55 & 19.273\\
         600 &  62 & 19.440\\
         800 &  66 & 19.566\\
         1000&  67 & 19.608\\
         1200&  75 & 19.620\\
        \hline
	\end{tabular}
	\label{sizeours}
\end{table}

\subsubsection{Comparison of Database Size}
As should be expected, we compare our method with other models based on mapping rules. Due to the different capacities utilized in these methods, it is impossible to make a direct comparison between their image database sizes. All methods are supposed to be expanded to 19 bits based on their original methods, and then the database size value can be attained via calculation. Considering that we could not get a sufficient amount of images to finish up the experiment, we had to make a manual estimation of the theoretical minimum. The results are shown in the Table \ref{sizecompare}, which show us  that other methods require an enormous amount of effort and time to construct an image database, whereas our method only necessitates around 200 random images to build the same. Consequently, our method has a great predominance in database management by comparison.

\begin{table*}[htpb]
	\centering
	\caption{Database size required for different methods (for 20-bit capacity)}
	\begin{tabular}{ccc}
		\hline
	Methods & Descriptions & Number of database images \\
        \hline
        ours & It generates a scrambling factor to hide information & 200 \\
        Mask RCNN \cite{2020Segment}& It generates $k$ binary sequence based on object areas & $\geq 174763$\\
        Multiple Objects \cite{2020Coverless}& It generates a sequence with fix length based on the number of objects & $\geq 524288$\\
        Faster RCNN \cite{2019Faster}& ditto & $\geq 524288$ \\
        Pixel \cite{2015Coverless}& It generates a sequence with fix length based on average pixel & $\geq 524288$\\
        SIFT  \cite{2017SIFT}& It generates a sequence with fix length based on SIFT feature & $\geq 524288$\\
        DCT \cite{2018Robust}& It generates $l$ binary sequence based on the DCT coefficients of the sub-blocks & $\geq 2048$ \\
        DWT \cite{2019Coverless}& It generates $l$ binary sequence based on the DWT coefficients of the sub-blocks & $\geq 2048$\\
        \hline
	\end{tabular}
	\label{sizecompare}
\end{table*}

\subsection{Robustness}
We use 4 common geometric attacks and 7 common noise attacks test the robustness of some typical methods under different parameters. And the rate of secret information extracted correctly is used to measure robustness. The calculation method for:
\begin{equation}\label{14}
    R=\frac{SM_c}{SM_o}
\end{equation}
Where $SM_c$ is the length of the correctly extracted secret information, and $SM_o$ is the length of the original information.

For geometric attacks, Table \ref{geometriccompare} shows that our method have a good promotion in robustness,  this is because the precision of the object detection model and the details of the mapping rules. Facing to cropping attacks, in particular, it has an ascending advantage over other deep learning-based methods. From Table \ref{noisecompare} , compared to noise attacks of traditional methods, our robustness is slightly weaker by reason of the limit to the accuracy of existing object detection models. But the performance is still impressive, with our method achieving an average of 87.4\%, which is marginally better than other deep learning models.

\begin{table*}[htbp]
	\centering
	\caption{The robustness of geometric attacks}
	\begin{tabular}{cccccccccc}
		\hline
		 Attack & parameters & Ours & Multiple objects \cite{2020Coverless}& Faster RCNN \cite{2019Faster}& Pixel \cite{2015Coverless}& SIFT \cite{2017SIFT}& DCT \cite{2018Robust}& DWT \cite{2019Coverless}\\
	    \hline
		 Center cropping & 5\% & 84.4\% & 51.2\% &  27.6\% & 47.4\% & 42.8\% & 48.4\% & 47.6\%  \\
                         &10\% & 76.0\% & 46.2\% &  23.2\% & 29.4\% & 22.6\% & 30.2\% & 27.0\% \\
                         &20\% & 57.3\% & 34.8\% &  16.4\% & 22.4\% & 6.0\%  & 22.6\% & 15.2\% \\
		 Edge cropping   &5\%  & 94.2\% & 87.6\% &  59.6\% & 58.2\% & 31.2\% & 59.4\% & 64.2\% \\
                         &10\% & 95.6\% & 86.2\% &  57.8\% & 38.8\% & 13.0\% & 39.8\% & 45.8\%  \\
                         &20\% & 90.2\% & 82.0\% &  54.0\% & 23.2\% & 6.8\%  & 21.6\% & 23.8\% \\
		 Rotation        &10   & 86.7\% & 78.4\% &  50.8\% &  8.0\% & 2.6\%  & 8.0\%  & 8.6\% \\
                         &30   & 61.3\% & 63.2\% &  40.2\% & 1.4\%  & 1.8\%  & 1.4\%  & 0.8\% \\
                         &50   & 50.7\% & 46.4\% &  29.8\% & 1.6\%  & 1.2\%  & 1.8\%  & 0.8\% \\
		 Translation   &(36,20)& 89.8\% & 83.2\% &  54.8\% & 20.4\% & 4.6\%  & 20.2\% & 17.2\% \\
                       &(40,25)& 87.6\% & 83.6\% &  54.8\% & 16.4\% & 3.8\%  & 16.6\% & 13.0\% \\
                       &(80,50)& 84.4\% & 77.0\% &  50.2\% & 6.0\%  & 2.0\%  & 5.2\%  & 4.8\% \\
        \hline
	\end{tabular}
	\label{geometriccompare}
\end{table*}

\begin{table*}[htbp]
	\centering
	\caption{The robustness of noise attacks}
	\begin{tabular}{c  c c cccccccc}
		\hline
		 Attack & parameters & Ours & Multiple objects \cite{2020Coverless}& Faster RCNN \cite{2019Faster}& Pixel \cite{2015Coverless}& SIFT \cite{2017SIFT}& DCT \cite{2018Robust}& DWT \cite{2019Coverless}\\
	    \hline
		 Gaussian noise         & 0.001 & 67.1\% & 71.8\% &  43.2\% & 95.8\% & 65.6\% & 95.4\% & 98.0\%  \\
    Salt and pepper noise       & 0.001 & 89.8\% & 86.0\% &  59.2\% & 99.0\% & 90.8\% & 99.2\% & 99.6\% \\
         Speckle noise          & 0.01  & 89.3\% & 83.6\% &  56.4\% & 96.6\% & 74.4\% & 96.2\% & 98.0\% \\
         Median filtering  &$3 \times 3$& 89.7\% & 87.2\% &  56.6\% & 99.6\% & 88.4\% & 99.4\% & 100\%  \\
         Mean filtering    &$3 \times 3$& 90.2\% & 86.0\% &  57.0\% & 98.8\% & 73.6\% & 95.8\% & 97.8\% \\
         Gaussian filtering&$3 \times 3$& 91.6\% & 89.6\% &  61.2\% & 99.8\% & 92.6\% & 100\%  & 100\% \\
              Scaling           &3      & 94.2\% & 91.2\% &  67.8\% & 99.6\% & 95.2\% & 100\%  & 100\% \\
        \hline
	\end{tabular}
	\label{noisecompare}
\end{table*}

\subsection{Security}
Our method effectively conceals secret information by forming a bond between objects, scrambling elements and a sequence key. The statistical properties of the image remain unchanged, meaning that existing steganalysis techniques cannot detect stego images that contain the secret information. Thus it can be seen that our method has an impressive level of resistance to steganalysis.

Even if the stego images are suspected by the attacker and the $Key$ is also cracked, the attacker would still be unable to extract the secret information, because the mapping dictionary and sequence key are agreed upon in advance and kept strictly confidential by both the sender and receiver. Despite the fact that one receiver inadvertently leaked the mapping dictionary and its sequence key, the security of the other receivers was not compromised, since each receiver has its own unique sequence key. This demonstrates the robustness of our method in terms of safety.

\section{Conclusions}
Our innovative coverless steganography relies on dynamic mapping, avoiding direct connections between images and binary sequences. A scrambling factor is extracted from the cover image, generating a new sequence for information hiding. Dynamic stego image selection maximizes capacity within sequence length constraints, effectively reducing the database load. Each image corresponds to numerous sequences with varying lengths, easing database construction and maintenance. Our method withstands all steganalysis, ensuring impeccable safety. Experimental results demonstrate robustness to common attacks. While our capacity currently lags behind traditional steganography, future work will aim to significantly enhance concealed information capabilities in coverless steganography.

\begin{small}
\bibliographystyle{ieeetr}
\bibliography{reference}

\begin{thebibliography}{10}

\bibitem{2008Adaptive}
C.~H. Yang, C.~Y. Weng, S.~J. Wang, and H.~M. Sun, ``Adaptive data hiding in
  edge areas of images with spatial lsb domain systems,'' {\em IEEE
  Transactions on Information Forensics \& Security}, vol.~3, no.~3,
  pp.~p.488--497, 2008.

\bibitem{2007Strange}
R.~T. Mckeon, ``Strange fourier steganography in movies,'' {\em IEEE
  International Conference on Electro/information Technology}, 2007.

\bibitem{2017A}
M.~Y. Valandar, P.~Ayubi, and M.~J. Barani, ``A new transform domain
  steganography based on modified logistic chaotic map for color images,'' {\em
  Journal of Information Security \& Applications}, vol.~34, 2017.

\bibitem{2002Secure}
I.~J. Cox, ``Secure spread spectrum watermarking for images, audio and video,''
  {\em IEEE Int.conf.image Processing}, vol.~3, 2002.

\bibitem{2008An}
W.~H. Lin, S.~J. Horng, T.~W. Kao, P.~Fan, C.~L. Lee, and Y.~Pan, ``An
  efficient watermarking method based on significant difference of wavelet
  coefficient quantization,'' {\em IEEE transactions on multimedia}, vol.~10,
  no.~5, pp.~p.746--757, 2008.

\bibitem{2015Coverless}
Z.~Zhou, H.~Sun, R.~Harit, X.~Chen, and X.~Sun, ``Coverless image steganography
  without embedding,'' {\em International Conference on Cloud Computing \&
  Security}, 2015.

\bibitem{2017SIFT}
S.~Zheng, W.~Liang, B.~Ling, and D.~Hu, ``Coverless information hiding based on
  robust image hashing,'' {\em International Conference on Intelligent
  Computing}, 2017.

\bibitem{2019A}
L.~Zou, J.~Sun, G.~Min, W.~Wan, and B.~B. Gupta, ``A novel coverless
  information hiding method based on the average pixel value of the
  sub-images,'' {\em Multimedia Tools and Applications}, 2019.

\bibitem{2018Robust}
X.~Zhang, F.~Peng, and M.~Long, ``Robust coverless image steganography based on
  dct and lda topic classification,'' {\em IEEE Transactions on Multimedia},
  pp.~3223--3238, 2018.

\bibitem{2019Coverless}
Q.~Liu, X.~Xiang, J.~Qin, Y.~Tan, and Y.~Luo, ``Coverless steganography based
  on image retrieval of densenet features and dwt sequence mapping,'' {\em
  Knowledge-Based Systems}, vol.~192, p.~105375, 2019.

\bibitem{2020Coverless}
Y.~Luo, J.~Qin, X.~Xiang, and Y.~Tan, ``Coverless image steganography based on
  multi-object recognition,'' {\em IEEE Transactions on Circuits and Systems
  for Video Technology}, vol.~PP, no.~99, pp.~1--1, 2020.

\bibitem{2020DenseNet}
Q.~Liu, X.~Xiang, J.~Qin, Y.~Tan, and Y.~Qiu, ``Coverless image steganography
  based on densenet feature mapping,'' {\em EURASIP Journal on Image and Video
  Processing}, vol.~2020, no.~1, p.~39, 2020.

\bibitem{2016Densely}
G.~Huang, Z.~Liu, V.~Laurens, and K.~Q. Weinberger, ``Densely connected
  convolutional networks,'' {\em IEEE Computer Society}, 2016.

\bibitem{2020Segment}
Y.~J. Luo, J.~Qin, X.~Xiang, Y.~Tan, and N.~N. Xiong, ``Coverless image
  steganography based on image segmentation,'' {\em Computers, Materials and
  Continua}, vol.~64, no.~2, pp.~1281--1295, 2020.

\bibitem{Camouflage}
Q.~Liu, X.~Xiang, J.~Qin, Y.~Tan, and Q.~Zhang, ``A robust coverless
  steganography scheme using camouflage image.,'' {\em IEEE Transactions on
  Circuits and Systems for Video Technology.}, 2021.

\bibitem{GAN}
I.~Goodfellow, J.~Pouget-Abadie, M.~Mirza, B.~Xu, D.~Warde-Farley, S.~Ozair,
  A.~Courville, and Y.~Bengio, ``Generative adversarial nets,'' {\em Neural
  Information Processing Systems}, 2014.

\bibitem{2017Coverless}
M.~M. Liu, M.~Q. Zhang, J.~Liu, Y.~N. Zhang, and Y.~Ke, ``Coverless information
  hiding based on generative adversarial networks,'' {\em Journal of Applied
  Sciences}, 2017.

\bibitem{2019An}
S.~Zhang, S.~Su, L.~Li, Q.~Zhou, and C.~C. Chang, ``An image style transfer
  network using multilevel noise encoding and its application in coverless
  steganography,'' {\em Symmetry}, vol.~11, no.~9, p.~1152, 2019.

\bibitem{2020A}
X.~Chen, Z.~Zhang, A.~Qiu, Z.~Xia, and N.~Xiong, ``A novel coverless
  steganography method based on image selection and stargan,'' {\em IEEE
  Transactions on Network Science and Engineering}, vol.~PP, no.~99, pp.~1--1,
  2020.

\bibitem{2004Distinctive}
D.~G. Lowe, ``Distinctive image features from scale-invariant keypoints,'' {\em
  International Journal of Computer Vision}, 2004.

\bibitem{0StarGAN}
Y.~Choi, M.~Choi, M.~Kim, J.~W. Ha, and J.~Choo, ``Stargan: Unified generative
  adversarial networks for multi-domain image-to-image translation,'' {\em 2018
  IEEE/CVF Conference on Computer Vision and Pattern Recognition (CVPR)}, 2018.

\bibitem{multi-domain}
W.~Y. Xue~R, ``Message drives image: A coverless image steganography framework
  using multi-domain image translation,'' {\em International Joint Conference
  on Neural Networks (IJCNN). IEEE.}, 2021.

\bibitem{GradientDescent}
L.~M. Peng~F, Chen~G, ``A robust coverless steganography based on generative
  adversarial networks and gradient descent approximation,'' {\em IEEE
  Transactions on Circuits and Systems for Video Technology.}, 2022.

\bibitem{2019Faster}
Z.~Zhou, Y.~Cao, and M.~Wang, ``Faster-rcnn based robust coverless information
  hiding system in cloud environment,'' {\em IEEE Access}, vol.~PP, no.~99,
  pp.~1--1, 2019.

\end{thebibliography}
\end{small}

\end{document}